\documentclass[format=acmsmall, review=false, screen=true]{acmart}
\acmJournal{TEAC}

\usepackage{booktabs} 

\usepackage[ruled]{algorithm2e} 
\usepackage{graphicx}
\graphicspath{ {./} }

\SetAlFnt{\small}
\SetAlCapFnt{\small}
\SetAlCapNameFnt{\small}
\SetAlCapHSkip{0pt}
\IncMargin{-\parindent}

\setcopyright{acmlicensed}

\acmDOI{}

\begin{document}
\title[Illustrating Greedy vs Dynamic Algorithms with Opportunity Cost]{Illustrating the Suitability of Greedy and Dynamic Algorithms Using Economics's ``Opportunity Cost''}
\author{Eugene Callahan}
\affiliation{%
  \institution{Tandon School of Engineering\\New York University}
}
\email{ejc369@nyu.edu}

\author{Robert Murphy}
\affiliation{%
  \institution{Free Market Institute\\Texas Tech University}
}
\email{robert.p.murphy@ttu.edu}

\author{Anas Elghafari}
\affiliation{%
  \institution{Tandon School of Engineering\\New York University}
}
\email{anas.elghafari@nyu.edu}

\begin{abstract}
\textbf{ABSTRACT\\}
Students of Computer Science often wonder when, exactly, one can apply a greedy algorithm to a problem, and when one must use the more complicated and time-consuming techniques of dynamic programming. This paper argues that the existing pedagogical literature does not offer clear guidance on this issue. We suggest improving computer science pedagogy by importing a concept economists use in their own implementations of dynamic programming. That economic concept is "opportunity cost," and we explain how it can aid students in differentiating "greedy problems" from problems requiring dynamic programming.\\
\end{abstract}

%
%


%
%


\acmYear{2020}
\acmMonth{11}
\maketitle

\renewcommand{\shortauthors}{Callahan, Elghafari, Murphy}

\section{Introduction}

Greedy algorithms simply seize upon the ``best'' choice available at every step in the algorithm.  When applied to suitable problems, this sequence of locally optimal choices will lead to the global optimum. Dynamic programming, on the other hand, tracks the impact of current choices on the chooser's ability to make other, future choices, ones that might lead to a global optimum but would be blocked by making the choice that ``seems best'' at any one moment. But when can one employ a (simpler and faster) greedy algorithm, and when must one resort to (slower and more complex) dynamic programming? The answer turns on the difference between static and dynamic optimization problems.\\

Specifically, in a static optimization problem, the actor is confronted with ``the same'' problem repeatedly, time and again. The choice in one period doesn't constrain the options in future periods, and so the overall global optimum (for the entire time horizon) coincides with the sequence of locally optimal decisions. In contrast, in a truly dynamic optimization problem, the choice in one period may significantly constrain the options available in future periods. Thus a decision in the present period--even though it may appear optimal, given its immediate consequences--may actually be a very poor decision, considered in the long run. Economists use the concept of opportunity cost to capture the idea that the true ``downside'' of a decision is the the potential payoff that is now rendered inaccessible because of the decision.\\

This paper aims to show that, while economics stresses the concept of opportunity cost, and offers a clear distinction between static and dynamic optimization problems, the distinction is not made as clearly in computer science. And what's more, we suggest that the computer science literature, especially the pedagogical literature, could benefit from incorporating this distinction. Here, for instance, is the way today's top algorithm textbook describes when a greedy algorithm may be used:

\begin{quote}
How can we tell whether a greedy algorithm will solve a particular optimization problem? No way works all the time, but the greedy-choice property and optimal substructure are the two key ingredients. If we can demonstrate that the problem has these properties, then we are well on our way to developing a greedy algorithm for it. (CLRS, p. 424)\cite{cormen2009introduction}\\
\end{quote}

To reiterate, the economists' idea of the opportunity cost of a choice can help clarify this situation. The opportunity cost of a choice is formally defined as the value that the chooser places on the next-best alternative that was (or will be) foregone in the making of the choice.(Gwartney et al. 2003, p. 10.)\cite{gwartney2003economics} For instance, we might be on the fence between going to Bangkok or to Paris for our vacation. If, in the end, we choose Bangkok, then the opportunity cost of the choice is the value we place on the vacation to Paris that we are now unable to take.\\

There are choices for which the opportunity cost manifests itself immediately at the moment of choice. For example, if we are  choosing between blueberry and apple pie, the cost we incur in choosing apple may be only the lack of pleasure we could have received from eating the slice of blueberry pie instead; this would be the case (perhaps) if we suppose the two choices require the same amount of money, contain the same number of calories, have the same amount of sugar, and so on.\\

On the other hand, if we are choosing between a slice of pie and a plate of broccoli, we may want to consider future effects. We may enjoy the pie more at the moment, but later regret its effect on our weight, or on our blood sugar levels. The opportunity cost of choosing the pie now is not just forgoing the immediate sensations associated with eating the broccoli, but also includes the weight gain and blood sugar effects that will only occur in a future "period."\\

The first example is one in which a greedy algorithm can be used: We simply choose the flavor of pie whose taste we prefer, since the future impact of either choice is identical. There is no need to consider other periods. On the other hand, when the choice has effects that spill over into future periods, as in the pie versus broccoli example, we must resort to dynamic programming.\\

\section{The Treatment of This Topic in the Computer Science Literature}
The concept of opportunity cost is not unknown in the computer science literature. For instance, Amir et al. \cite{amir2000opportunity} and Borgstrom (2000)\cite{borgstrom2001cost} employ it to analyze job scheduling problems in a metacomputer. Pillac et al. (2011) \cite{pillac2011dynamic} employ the concept in analyzing vehicle routing problems. Nevertheless, we have not been able to find an instance in the computer science literature of opportunity cost being used to guide the choice between a greedy approach and a dynamic programming approach. Lew comes close to stating the economist's distinction, but with less clarity:

\begin{quote}
Informally, a ``greedy'' optimization algorithm solves a global minimization or maximization problem by making a sequence of locally optimal decisions. A decision is ``local'' or ``myopic'' if it is based on partial information that does not include global knowledge about future consequences of current decisions (Heyman and Sobel, 1984); thus, the current decision, once made, might turn out to be nonoptimal. (Lew 2006, p. 621)\cite{lew2006canonical}
\end{quote}

Lew later states:

\begin{quote}
While we lack a general procedure to determine whether or not a canonical greedy algorithm is optimal, no general procedure exists for noncanonical greedy algorithms either.
\end{quote}

\section{The Treatment of This Topic in the Economic Literature}
Ferguson and Lim \cite{ferguson2005discrete} capture this idea in their distinction between static and dynamic consumption problems:
\begin{quote}
In a static consumption problem the opportunity cost of spending an amount of money to buy a unit of one commodity (and deriving the marginal utility associated with consuming one more unit of that commodity) is the largest extra utility that would have been derived from spending money some other on some other commodity. In a dynamic problem the opportunity cost of spending today is the largest extra lifetime utility we could have derived from saving the money and spending it at some point in the future.
\end{quote}

In their terms, when we have a static consumption problem (i.e., the opportunity cost is confined to the current period), we can use a greedy algorithm. When the opportunity cost must be considered over future periods, we must resort to a more complicated technique, such as dynamic programming\\

We can illustrate this distinction with two simple examples, which are both typical optimization problems of the kind students might encounter in an economics graduate program. First, suppose a manager must decide how many units of output $Y_t$ to produce each period $t$, from $t=1$ to $N$, given the market price $P_t$ of the output at time $t$, and where the total cost in period t is a function of the level of output and is given by $C_t (Y_t )= 1000 + Y_t^2$ \\

In an economics problem of this sort, the only ``global check'' that we would have to perform, is to make sure that the manager wants to keep the plant in operation at all (and keep suffering the fixed cost of 1000 each period). But, assuming that the plant stays in operation, then solving the ``myopic'' or ``greedy'' optimization period by period, will also yield the long-run or global optimum.\\

In contrast, consider a different type of problem--which is, again, typical in a graduate economics program. Suppose a household has a stock of capital goods each period, $K_t$, that it uses in a production function $f$ to produce output according to $Y_t=f(K_t)$. Now each period the household must decide how much of the new output to consume as $C_t$, and how much to devote to gross investment and augment the capital stock to be carried forward to next period. However, there is also physical depreciation by a factor of $0 < \delta < 1$, meaning the proportion of the capital stock that is used up each period during production. When the household consumes, it receives an immediate burst of utility according to the function $u(C_t)$. However, from the perspective of period t, a burst of utility accruing next period will be discounted and only register as $\beta u(C_{t+1})$, where $0 < \beta < 1$.\\

Given this setup, the household wants to optimize its lifetime flow of utility by choosing the optimal path of consumption in each discrete time period. In principle this is a difficult problem, but standard techniques on recursive problems (see e.g. Stokey and Lucas 1989 \cite{stokey1989recursive}) show how to set up a ``Bellman equation'' using a value function:
$$V(K_t) = \max\{u(C_t) +\beta V(K_{t+1})\}$$
such that:
$$K_{t+1} = f(K_t) - C_t + (1 -  \delta) K_t    \footnote{
Both to ensure a well-defined problem and to match the math with the economic intuition, such problems would carry further assumptions on the functional forms, for example insisting that the first derivatives of $f()$ and $u()$ were positive while the second derivatives were negative, and that the limit of the first derivative of $u(x)$ as $x\rightarrow 0$ is infinity. We are neglecting an exhaustive treatment of these subtleties for the present discussion, because the important thing is to show how the economics literature determines when to formally use dynamic programming.}$$

The intuition behind the Bellman equation is that the value in period $t$ of holding $K_t$ of capital, is the present discounted flow of future bursts of utility, assuming that the household chooses optimally this period and in all future periods. But by definition, $V(K_{t+1})$ is the value next period for the household at that point, given that it has inherited $K_{t+1}$ capital stock from period $t$. Thus, the term inside the $max$ operator captures the opportunity cost involved: increasing $C_t$ this period boosts immediate gratification in terms of the utility function, but because of the constraint it reduces the amount of capital passed on to the next period.\\

In this type of problem, it would be obviously foolish to maximize utility in the present period, disregarding the future. Therefore, it is the epitome of a problem conducive to dynamic programming methods for the solution.\\

\section{Application to Choosing between Greedy and Dynamic Algorithms}
To show the explanatory power of the concept of opportunity cost, contrast two graph problems: finding the longest (or maximum-benefit) path between two nodes, versus finding the maximum spanning tree. We know that a greedy approach is not guaranteed to find the maximum-benefit path but is guaranteed to find the maximum spanning tree. So, can we motivate this distinction using the notion of opportunity cost?
\begin{figure}[h]
\centering
\includegraphics[scale=0.65]{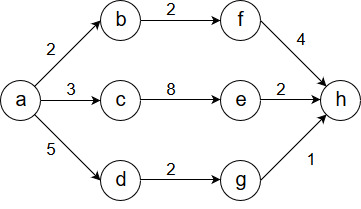}
\caption{Example Graph}
\label{graph1}
\end{figure}

\subsection{Dynamic Programming case: Finding Maximum-Benefit Path}
Without loss of generality, we will use \ref{graph1} for illustration. We will define the opportunity cost of choosing an edge $x-y$ to be the utility of the best forgone path. So, in this particular example where 3 distinct paths exist from $a$ to $h$, choosing edge $a-d$ means forgoing two paths: $a-b-f-h$ and $a-c-e-h$. So, the opportunity cost of this choice is the maximum utility of these 2 paths $OPPCOST(a-d)= MAX\{u(a-b-f-h),u(a-c-e-h)\}$\\

Given the task of finding the maximum utility path from a to h, the 3 initial choices are: $a-b,a-c,a-d$, at utilities 2,3, and 5, respectively. A greedy approach for this task would choose $a-d$ since it is the edge with the maximum utility, and a student who is new to this topic will not immediately see why this is not an optimal choice. However, the sub-optimality of this choice can be illustrated by prompting the student to consider not only the utility of each edge, but the opportunity cost as well. \\

$$OPPCOST(a-b)= MAX\{u(a-c-e-h), u(a-d-g-h)\} = u(a-c-e-h) = 13$$
$$OPPCOST(a-c)= MAX\{u(a-b-f-h),u(a-d-g-h)\} = u(a-b-f-h) = 8$$
$$OPPCOST(a-d)= MAX\{u(a-b-f-h),u(a-c-e-h)\} = u(a-c-e-h) = 13$$\\

In this case, edge $a-d$ has the opportunity cost of the utility of the better of the two forgone paths: $a-b-f-h$ and $a-c-e-h$, which is 13. Similarly, the opportunity cost of choosing $a-b$ is also 13. However, considering the opportunity cost of $a-c$ reveals something interesting: the utility of the best forgone path is 8. This is 5 units lower than $OPPCOST(a-d)$, the greedy choice. This demonstrates that taking the greedy option, while it yields 2 units higher in immediate utility, will cause us to miss out on 5 units of utility. As such, the greedy local choice is a worse choice overall.\\

\subsection{Greedy case: Finding the Maximum-Spanning Tree}
Let us now consider a task where a greedy approach yields the optimal solution and see how the suitability of the greedy approach can be grounded in the language of opportunity cost. \\

Recall that a spanning tree of a graph G is a tree that includes all vertices of G, and that a spanning tree is said to be ``maximum'' if no other spanning tree has a higher sum for the weights of its edges. A maximum spanning tree for a graph can be found using Kruskal's algorithm: given a graph G, we order the set of edges by descending weight and continually add the highest-weighted edge connecting two vertices that are not yet connected in our tree. We stop when a spanning tree is obtained. In this greedy algorithm it is guaranteed that the resulting spanning tree will be maximum. We will show how the concept introduced above can be used to illustrate this case, where the greedy choice coincides with the optimal one.\\

In applying Kruskal to the example graph above, the ordering of the edges we obtain is \\\\
$[c-e: 8,a-d: 5,f-h:4,a-c:3,a-b: 2,b-f: 2,d-g: 2,e-h:2,g-h:1]$.\\
Applying the same definition for opportunity cost as above (max utility of forgone path) we get that $OPPCOST(c-e) = 5$, while 
$OPPCOST(a-d) = 8$, $OPPCOST(f-h) = 8$, and so on. Since the choice of an edge does not restrict future choices, the greedy choice will not have a higher opportunity cost than any non-greedy choice. We can see that the greedy choice always has the lowest opportunity cost, demonstrating that a greedy algorithm applied to this task will yield the optimal solution.\\

\section{Conclusion}
The practice of dynamic programming arose in the context of operational planning, where the idea of opportunity costs was generally understood. But as computer scientists began to treat the topic in algorithms textbooks and elsewhere, the economic context of these sorts of problems receded into the background, while pure algorithmic analysis came to the fore. Unfortunately, in the process, an important concept--that of opportunity cost--was lost, even though it is of great help in distinguishing cases where dynamic programming must be employed versus ones in which a greedy algorithm is sufficient. We recommend re-introducing the concept of opportunity cost into the computer science literature, as we believe it has great pedagogical value.\\\\

\bibliographystyle{ACM-Reference-Format}
\bibliography{sample-bibliography}


\begin{thebibliography}{8}


\ifx \showCODEN    \undefined \def \showCODEN     #1{\unskip}     \fi
\ifx \showDOI      \undefined \def \showDOI       #1{#1}\fi
\ifx \showISBNx    \undefined \def \showISBNx     #1{\unskip}     \fi
\ifx \showISBNxiii \undefined \def \showISBNxiii  #1{\unskip}     \fi
\ifx \showISSN     \undefined \def \showISSN      #1{\unskip}     \fi
\ifx \showLCCN     \undefined \def \showLCCN      #1{\unskip}     \fi
\ifx \shownote     \undefined \def \shownote      #1{#1}          \fi
\ifx \showarticletitle \undefined \def \showarticletitle #1{#1}   \fi
\ifx \showURL      \undefined \def \showURL       {\relax}        \fi
\providecommand\bibfield[2]{#2}
\providecommand\bibinfo[2]{#2}
\providecommand\natexlab[1]{#1}
\providecommand\showeprint[2][]{arXiv:#2}

\bibitem[\protect\citeauthoryear{Amir, Awerbuch, Barak, Borgstrom, and
  Keren}{Amir et~al\mbox{.}}{2000}]%
        {amir2000opportunity}
\bibfield{author}{\bibinfo{person}{Yair Amir}, \bibinfo{person}{Baruch
  Awerbuch}, \bibinfo{person}{Amnon Barak}, \bibinfo{person}{R~Sean Borgstrom},
  {and} \bibinfo{person}{Arie Keren}.} \bibinfo{year}{2000}\natexlab{}.
\newblock \showarticletitle{An opportunity cost approach for job assignment in
  a scalable computing cluster}.
\newblock \bibinfo{journal}{\emph{IEEE Transactions on parallel and distributed
  Systems}} \bibinfo{volume}{11}, \bibinfo{number}{7} (\bibinfo{year}{2000}),
  \bibinfo{pages}{760--768}.
\newblock


\bibitem[\protect\citeauthoryear{Borgstrom, Awerbuch, and Amir}{Borgstrom
  et~al\mbox{.}}{2001}]%
        {borgstrom2001cost}
\bibfield{author}{\bibinfo{person}{R~Sean Borgstrom}, \bibinfo{person}{Baruch
  Awerbuch}, {and} \bibinfo{person}{Yair Amir}.}
  \bibinfo{year}{2001}\natexlab{}.
\newblock \bibinfo{booktitle}{\emph{A cost-benefit approach to resource
  allocation in scalable metacomputers}}.
\newblock \bibinfo{type}{{T}echnical {R}eport}. \bibinfo{institution}{JOHNS
  HOPKINS UNIV BALTIMORE MD}.
\newblock


\bibitem[\protect\citeauthoryear{Cormen, Leiserson, Rivest, and Stein}{Cormen
  et~al\mbox{.}}{2009}]%
        {cormen2009introduction}
\bibfield{author}{\bibinfo{person}{Thomas~H Cormen}, \bibinfo{person}{Charles~E
  Leiserson}, \bibinfo{person}{Ronald~L Rivest}, {and}
  \bibinfo{person}{Clifford Stein}.} \bibinfo{year}{2009}\natexlab{}.
\newblock \bibinfo{booktitle}{\emph{Introduction to algorithms}}.
\newblock \bibinfo{publisher}{MIT press}.
\newblock


\bibitem[\protect\citeauthoryear{Ferguson and Lim}{Ferguson and Lim}{2005}]%
        {ferguson2005discrete}
\bibfield{author}{\bibinfo{person}{Brian Ferguson} {and} \bibinfo{person}{Guay
  Lim}.} \bibinfo{year}{2005}\natexlab{}.
\newblock \bibinfo{booktitle}{\emph{Discrete time dynamic economic models:
  Theory and empirical applications}}.
\newblock \bibinfo{publisher}{Routledge}.
\newblock


\bibitem[\protect\citeauthoryear{Gwartney, Stroup, Sobel, and
  Macpherson}{Gwartney et~al\mbox{.}}{2003}]%
        {gwartney2003economics}
\bibfield{author}{\bibinfo{person}{James~D Gwartney},
  \bibinfo{person}{Richard~L Stroup}, \bibinfo{person}{S~Russell Sobel}, {and}
  \bibinfo{person}{David~A Macpherson}.} \bibinfo{year}{2003}\natexlab{}.
\newblock \showarticletitle{Economics: private and public choice. 10 Th}.
\newblock \bibinfo{journal}{\emph{Editions. Thomson, South Western}}
  (\bibinfo{year}{2003}).
\newblock


\bibitem[\protect\citeauthoryear{Lew}{Lew}{2006}]%
        {lew2006canonical}
\bibfield{author}{\bibinfo{person}{Art Lew}.} \bibinfo{year}{2006}\natexlab{}.
\newblock \showarticletitle{Canonical greedy algorithms and dynamic
  programming}.
\newblock \bibinfo{journal}{\emph{Control and Cybernetics}}
  \bibinfo{volume}{35}, \bibinfo{number}{3} (\bibinfo{year}{2006}),
  \bibinfo{pages}{621--643}.
\newblock


\bibitem[\protect\citeauthoryear{Pillac, Gu{\'e}ret, and Medaglia}{Pillac
  et~al\mbox{.}}{2011}]%
        {pillac2011dynamic}
\bibfield{author}{\bibinfo{person}{Victor Pillac}, \bibinfo{person}{Christelle
  Gu{\'e}ret}, {and} \bibinfo{person}{Andr{\'e}s Medaglia}.}
  \bibinfo{year}{2011}\natexlab{}.
\newblock \showarticletitle{Dynamic vehicle routing problems: State of the art
  and prospects}.
\newblock  (\bibinfo{year}{2011}).
\newblock


\bibitem[\protect\citeauthoryear{Stokey}{Stokey}{1989}]%
        {stokey1989recursive}
\bibfield{author}{\bibinfo{person}{Nancy~L Stokey}.}
  \bibinfo{year}{1989}\natexlab{}.
\newblock \bibinfo{booktitle}{\emph{Recursive methods in economic dynamics}}.
\newblock \bibinfo{publisher}{Harvard University Press}.
\newblock


\end{thebibliography}

\end{document}